\documentclass{PoS}
\usepackage{epsfig,amsmath}
\DeclareMathOperator{\im}{Im}
\DeclareMathOperator{\re}{Re}
\title{Dispersive construction of two-loop $P\rightarrow3\pi$ $(P=K,\eta)$ amplitudes}

\ShortTitle{Dispersive construction of two-loop $P\rightarrow3\pi$ $(P=K,\eta)$ amplitudes}

\author{\speaker{Martin ZDRAHAL}%
         \thanks{This work was supported in part by the Center for Particle Physics
         (project no.\ LC 527), GACR (grant no.\ 202/07/P249) and by the EU Contract
         No.\ MRTN-CT-2006-035482, \lq\lq{\sc Flavia\it net}''.}\\
        Faculty of Physics, University of Vienna, Boltzmanngasse 5, A-1090 Vienna, Austria\\
        Faculty of Mathematics and Physics, Charles University, V~Hole\v{s}ovi\v{c}k\'{a}ch 2, Prague, Czech Rep.\\
        E-mail: \email{martin.zdrahal@univie.ac.at}}
\author{Karol Kampf\\Paul Scherrer Institut, CH-5232 Villigen PSI, Switzerland,\\
        Faculty of Mathematics and Physics, Charles University, V~Hole\v{s}ovi\v{c}k\'{a}ch 2, Prague, Czech Rep.\\
        E-mail: \email{karol.kampf@psi.ch}%
        }
\author{Marc Knecht\\Centre de Physique Th\'eorique\thanks{Unit\'e Mixte de Recherche (UMR 6207)
 du CNRS et des Universit\'es Aix--Marseille 1, Aix--Marseille 2 et du Sud
Toulon--Var, Laboratoire affili\'e \`a la FRUMAM (FR 2291)}
, CNRS-Luminy, Case 907, F-13288 Marseille Cedex 9, France\\
        E-mail: \email{marc.knecht@cpt.univ-mrs.fr}%
}
\author{Ji\v{r}\'i Novotn\'y\\
        Faculty of Mathematics and Physics, Charles University, V~Hole\v{s}ovi\v{c}k\'{a}ch 2, Prague, Czech Rep.\\
        E-mail: \email{novotny@ipnp.troja.mff.cuni.cz}%
}
\abstract{The branching ratio of the $\eta\to 3\pi$ decay is an important source of information on the value of the quark mass ratio $\frac{1}{R}=\frac{m_d-m_u}{m_s - {\widehat m}}$. Furthermore, isospin breaking effects in the decays $K\to 3 \pi$ provide information on the pion scattering lengths. The cusp effect in the $K\to 3 \pi$ decays is presently being analyzed by the NA48 and KTeV experiments. From the theoretical point of view, these processes have been studied by different methods. We propose a unified and relativistic treatment relying on very general principles, unitarity, analyticity and crossing symmetry, combined with chiral counting, in order to construct model-independent representations of the corresponding amplitudes that are valid at two loops. A general description of the procedure is given and is illustrated in the case of the $\eta$ decay amplitude in the leading order in the isospin breaking.}

\FullConference{International Workshop on Effective Field Theories: from the pion to the upsilon\\
         2-6 February 2009\\
         Valencia, Spain}

\renewcommand{\logo}
{\raisebox{-5mm}
{\parbox{\textwidth}{\begin{flushright}
CPT-P038-2009\\
PSI-PR-09-07\\
UWThPh-2009-04
\end{flushright}
\includegraphics{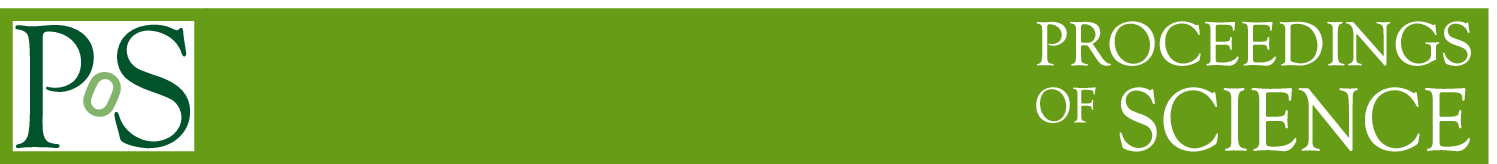}}
}}

\begin{document}
\section{Introduction}
During the last few years, the decay processes $K\rightarrow3\pi$ and $\eta\rightarrow3\pi$ have been under intensive studies, both from the experimental \cite{etacharged}--\cite{kaon} and from the theoretical \cite{kaontheory}--\cite{BN} points of view. The importance of these processes, besides the usual determination of decay rates and energy distributions, lies in the possibility of studying isospin breaking effects. The appearance of the cusp effect in those processes with two neutral pions in the final state enables quite a simple determination of the $\pi\pi$ scattering lengths (mainly from $K^+\rightarrow\pi^+\pi^0\pi^0$ decay). The $\eta$ decays, which are forbidden in the isospin limit, offer a good possibility to determine the isospin breaking parameters like $R$ and $Q$ from \cite{Bijnens}.

We present a method how to obtain a model-independent parametrization of the amplitudes of the decays
\begin{subequations}\label{procesy}
\begin{align}
K^+&\rightarrow\pi^+\pi^0\pi^0,\ \pi^+\pi^+\pi^-;\\
K_L&\rightarrow\pi^0\pi^0\pi^0,\ \pi^0\pi^+\pi^-;\\
\eta&\rightarrow\pi^0\pi^0\pi^0,\ \pi^0\pi^+\pi^-;\\
K_S&\rightarrow\pi^0\pi^+\pi^- .
\end{align}
\end{subequations}
This parameterization hinges on a two-loop construction of these amplitudes based only on general properties like unitarity, analyticity, crossing symmetry, relativistic invariance, and chiral power-counting for partial wave amplitudes. Up to two loops it takes the form
\begin{equation}\label{tvar amplitudy}
\mathcal{A}(s,t,u)=\mathcal{N}_F\left[\mathcal{P}(s,t,u)+\mathcal{U}(s,t,u)\right]+O(p^8),
\end{equation}
where $\mathcal{N}_F$ is an overall normalization, while the polynomial part $\mathcal{P}(s,t,u)$ contains free parameters describing the energy dependence of the processes in analogy to the Dalitz parameters of the traditional PDG parameterization. All the non-analytic part of the amplitude connected with the final state $\pi\pi$ scattering (as discussed below, we could include also other intermediate states but for our purpose the $\pi\pi$ ones are enough) is contained in $\mathcal{U}(s,t,u)$. It depends on the parameters from the polynomial part and on parameters describing the $\pi\pi$ scattering, like the scattering lengths, which in turn allows for their determination from the experimental study of the cusp effects.

In the following we shall concentrate on the $\eta$ decay, where the important physics appears already in the first order in isospin breaking (IB$^1$), mainly because of the less involved analytic expressions. Naturally, in this limit we cannot describe the cusp effect. The discussion of it is postponed to our forthcoming paper \cite{plan}, cf. with our previous proceeding \cite{proceed} where also a small summary of the literature concerning it is given.

The existing theoretical approaches to computing the $\eta$ decay amplitudes can be divided into several groups. The first one (mainly \cite{GL} and \cite{BG}) encompasses computation using chiral perturbation theory (ChPT), where the complete two-loop amplitudes at IB$^1$ are obtained. Nowadays, the most advanced approach (mainly in the vicinity of the cusp region) describing also the cusp effect and already including also the electromagnetic corrections is a method using the non-relativistic effective field theory \cite{Bern}. The third group includes the use of analyticity and unitarity for the construction of the amplitude. In 1996 there appeared almost at the same time two such methods trying to find the fixed point solution of Khuri-Treiman type dispersive relations. Kambor {\it et al.~}\cite{KWW} and independently Anisowich and Leutwyler \cite{AL} solved numerically this integral equation and then obtained values of the subtraction constants from the matching with ChPT at specific points. These approaches perform a numerical resummation of two-pion rescattering contributions. In contrast, our method provides an analytic dispersive representation valid to two loops that follows from general properties like unitarity, analyticity and crossing symmetry, combined with chiral counting. For the sake of completeness, let us add that there exist still other approaches using different methods, like the application of the Bethe-Salpeter equation to unitarised ChPT \cite{BN}.

\section{Reconstruction theorem}
As already sketched in our previous proceeding \cite{proceed}, we iteratively construct in parallel the $\pi\pi$ scattering amplitude and the amplitude of $P\pi\rightarrow\pi\pi$, related to the $P\rightarrow3\pi$ by crossing symmetry, using the same procedure as in \cite{SSFKMF,ZN}.

Provided we have the following chiral behaviour of the partial waves of the amplitude
\begin{gather}
\mathcal{A}(s,t,u)=16\pi\,\mathcal{N}_F(f_0(s)+3f_1(s)\cos \theta )+\mathcal{A}_{\ell\geq 2},\\
\re f_{\ell=0,1}(s) \sim O(p^2) ,\quad \im f_{\ell=0,1}(s) \sim O(p^4),\\
\re \mathcal{A}_{\ell\geq 2}\sim O(p^4),\quad \im \mathcal{A}_{\ell\geq 2}\sim O(p^8),
\end{gather}
we can reconstruct the amplitude in the form (\ref{tvar amplitudy}), with $\mathcal{P}(s,t,u)$ being a third order polynomial in the Mandelstam variables having the same $s,t,u$ symmetries as the amplitude $\mathcal{A}(s,t,u)$. The unitarity part
\begin{equation}
\mathcal{U}(s,t,u)=W_S(s,t,u)+W_T(t,s,u)+W_U(u,t,s)
\end{equation}
is given in terms of single variable dispersive integrals over the imaginary parts of S and P partial waves of all the crossed amplitudes.
For instance, the contribution of the s-channel amplitude $AB\rightarrow CD$ is
\begin{multline}\label{rekonstrukce}
W_S(s,t,u)=16s^3\int_{\text{thr}}^{\Lambda^2} \!\!dx\,\frac{1}{%
x^3(x-s)}\Big[\im f_0(x) + 3 \im f_1(x) \frac{(m_A^2-m_B^2)(m_C^2-m_D^2)}{\lambda _{AB}^{1/2}(x)\lambda _{CD}^{1/2}(x)}\Big]\\
+ 48 s^2(t-u)\int_{\text{thr}}^{\Lambda^2} \!\!dx\,\frac{1}{%
x(x-s)}\frac{\im f_1(x)}{\lambda _{AB}^{1/2}(x)\lambda _{CD}^{1/2}(x)}
\end{multline}
and similar for the $t$- and $u$-crossed-channel contributions $W_T$ and $W_U$ respectively. The triangle function $\lambda_{AB}(x)$ is defined as $\lambda_{AB}(x)=(x-(m_A+m_B)^2)(x-(m_A-m_B)^2)$.

To obtain the imaginary parts entering the above expressions, we use the unitarity relation projected on the corresponding partial waves,
\begin{equation}
\im f_\ell^{i\rightarrow f}(s)=\sum_k\frac{1}{S_k}\frac{\lambda^{1/2}_k(s)}{s} f_\ell^{i\rightarrow k}(s)\left(f_\ell^{f\rightarrow
k}(s)\right)^{*}\theta(s-\text{thr}_k).
\end{equation}
The sum goes over all the possible intermediate states $k$ with symmetry factor $S_k$ (in the case of two-particle states $k$, $S_k=2$ for undistinguishable states and $S_k=1$ otherwise) and $\text{thr}_k$ the threshold above which this channel opens. In the low-energy region and up to two loops, $k$ are restricted to be pairs of light pseudoscalar mesons. In the decay region these can be further restricted to intermediate $\pi\pi$ states only. The contributions from other intermediate states, like \emph{e.g.}\ $K\pi$, can be expanded in powers of the Mandelstam variables and absorbed into the polynomial $\mathcal{P}(s,t,u)$.

Thanks to this restriction on the relevant intermediate states, we can proceed iteratively as is shown in the following section on the case of the $\eta\rightarrow\pi^0\pi^+\pi^-$ in IB$^1$. But the argumentation goes similarly also for the other decays (\ref{procesy}) and the isospin symmetry plays there no important role \cite{ZN} other than that it simplifies the analytic form of the relations.

\section{Iterative construction of $\eta\rightarrow3\pi$ process}
In the isospin limit the three-pion decay of $\eta$ is forbidden. We shall construct the amplitude of this process at first order in isospin breaking, \emph{i.e.} we shall take the leading order amplitude of this decay nonzero, $\mathcal{N}_F\neq0$, and elsewhere in the calculation we will take an exact isospin limit.

In this limit the different amplitudes are related; the pion-scattering ones (in our sign convention) by
\begin{align}
\mathcal{A}_c(s,t,u)&=-\mathcal{A}_x(s,t,u)-\mathcal{A}_x(t,s,u),\\
\mathcal{A}_{00}(s,t,u)&=-\mathcal{A}_x(s,t,u)-\mathcal{A}_x(t,s,u)-\mathcal{A}_x(u,t,s)
\end{align}
with the notation
\begin{equation}
\mathcal{A}_{00}:\ \pi^0\pi^0\rightarrow \pi^0\pi^0,
\ \mathcal{A}_{x}:\ \pi^0\pi^0\rightarrow \pi^+\pi^-,
\ \mathcal{A}_{c}:\ \pi^+\pi^-\rightarrow \pi^+\pi^-;
\end{equation}
and the $\eta$ decay (or $\eta\pi\rightarrow\pi\pi$) amplitudes by
\begin{equation}
\tilde{\mathcal{A}}_{00}(s,t,u)=-\tilde{\mathcal{A}}_x(s,t,u)-\tilde{\mathcal{A}}_x(t,s,u)-\tilde{\mathcal{A}}_x(u,t,s),
\end{equation}
where
\begin{equation}
\tilde{\mathcal{A}}_{00}:\ \ \eta\pi^0\rightarrow \pi^0\pi^0,
\ \tilde{\mathcal{A}}_{x}:\ \eta\pi^0\rightarrow \pi^+\pi^-.
\end{equation}
It is therefore enough to reconstruct the $\mathcal{A}_x$ (reproducing the original Stern {\it et al.}\ computation \cite{SSFKMF}) and
$\tilde{\mathcal{A}}_x$ amplitudes.

\subsection{Leading order amplitudes}
From ChPT we know that at $O(p^2)$ both amplitudes are represented by first-order polynomials in the
Mandelstam variables. Their particular choice (connected also with the particular choice of the polynomial of the reconstruction
theorem) is important since different choices can possibly lead to different convergence properties of the chiral expansion and affect
the stability of the fit to the data. Contrary to \cite{proceed} we show here the parametrization of $\mathcal{A}_x$ using its
scattering length and effective range parameter. The parametrization of $\tilde{\mathcal{A}}_x$ is chosen using the subthreshold parameters.
The leading-order amplitudes then look like
\begin{align}
\mathcal{A}_{x}(s,t,u)&=16\pi\Big(\mathrm{a}+\frac{\mathrm{b}}{F_\pi^2}(s-4M_\pi^2)\Big)+O(p^4),\\
\tilde{\mathcal{A}}_{x}(s,t,u)&=\mathcal{N}_F\left(\tilde{A}M_\eta^2+\tilde{B}(s-s_0)\right)+O(p^4),
\end{align}
where $s_0=(M_\eta^2 + 3M_\pi^2)/3$ denotes the center of the Dalitz plot.

The $O(p^2)$ chiral perturbation theory result is reproduced by special values of the parameters. To get the pion scattering
amplitude in terms of the isospin scattering lengths \cite{GL1}, we can use $\mathrm{a}=\frac{1}{6}(a^2-a^0)$ and
$\mathrm{b}=\frac{F_\pi^2}{48M_\pi^2}(5a^2-2a^0)$. The $\eta\pi$ constants would be $\mathcal{N}_F\tilde{A}M_\eta^2=\frac{B_0(m_u-m_d)}{3\sqrt3
F_\pi^2}$ and $\mathcal{N}_F\tilde{B}=\frac{B_0(m_u-m_d)}{\sqrt3 F_\pi^2}\frac{1}{(M_\eta^2-M_\pi^2)} $ (see \cite{BG}).

\subsection{First iteration: one-loop expressions}
Using the partial waves computed from these amplitudes inside the reconstruction theorem, we obtain the $O(p^4)$ form of the $\pi\pi$
scattering amplitude
\begin{equation}
\mathcal{A}_x(s,t,u)=16\pi\,\big(\!\mathcal{P}_x(s,t,u)+W_x(s,t,u)+W_{+0}(t,s,u)+W_{+0}(u,s,t)\big)+O(p^6),
\end{equation}
where
\begin{align}
W_x(s,t,u)&=-8\pi \bar{J}(s)\Big(\mathrm{a}+\frac{\mathrm{b}}{F_\pi^2}(s-4M_\pi^2)\Big)\Big(7\mathrm{a}+\frac{\mathrm{b}}{F_\pi^2}(s-20M_\pi^2)\Big),\\
W_{+0}(s,t,u)&=-16\pi \bar
J(s)\bigg(\Big(\mathrm{a}-\frac{\mathrm{b}}{2F_\pi^2}(s+4M_\pi^2)\Big)^2+\frac{\mathrm{b}\,{}^2}{12F_\pi^4}(s-4M_\pi^2)(t-u)\bigg),\\
\bar J(s)&= \frac{1}{16\pi^2} \Big(2+\sigma(s) \ln \frac{\sigma(s) -1}{\sigma(s)+1}\Big),\\
\sigma(s)&=\sqrt{1-\frac{4M_\pi^2}{s}}
\end{align}
and $\mathcal{P}_x(s,t,u)$ is a second order polynomial in $s$, $t$ and $u$, symmetric in $t-u$ exchange. We can write the polynomial in
the form
\begin{equation}
\mathcal{P}_x(s,t,u)=\hat{\mathrm{a}}+\frac{\hat{\mathrm{b}}}{F_\pi^2}(s-4M_\pi^2)+w_x(s)-\frac{\lambda^{(1)}}{F_\pi^4}s(s-4M_\pi^2)
-\frac{\lambda^{(2)}}{F_\pi^4}\left[t(t-4M_\pi^2)+u(u-4M_\pi^2)\right],
\end{equation}
where
\begin{equation}
w_x(s)=\frac{\mathrm{a}}{\pi}\Big(7\mathrm{a}-16M_\pi^2\frac{\mathrm{b}}{F_\pi^2}\Big)-
\frac{s-4M_\pi^2}{36\pi
M_\pi^2}\Big(69\,\mathrm{a}^2-456M_\pi^2\,\mathrm{a}\frac{\mathrm{b}}{F_\pi^2}+592M_\pi^4\frac{\mathrm{b}^2}{F_\pi^4}\Big)
\end{equation}
is the polynomial canceling the real part of the S wave of the unitarity part near the threshold. Therefore, the scattering length and the effective
range parameter with hats keep their physical interpretation up to one loop. Motivated by this interpretation we can resum a part of the terms from the two loop chiral order and change also the non-hatted variables into their physical values up to one loop. Similarly, we could reconstruct the pion amplitudes up to two-loop order in a way where these coefficients keep their physical interpretation up to two loops, just as in the non-relativistic approach \cite{Bern}.

Using the same procedure, we can reconstruct also the $\eta\pi^0\rightarrow\pi^+\pi^-$ amplitude,
\begin{equation}\label{vysledek}
\tilde{\mathcal{A}}_x(s,t,u)=\mathcal{N}_F\big(\tilde{\mathcal{P}}_x(s,t,u)+\tilde{W}_x(s,t,u)+\tilde{W}_{+0}(t,s,u)+\tilde{W}_{+0}(u,s,t)\big)+O(p^6),
\end{equation}
with
\begin{align}
\tilde{W}_x(s,t,u)&=-8\pi \bar
J(s)\bigg(\tilde{A}M_\eta^2\Big(7\mathrm{a}+4\frac{\mathrm{b}}{F_\pi^2}(s-6M_\pi^2)\Big)+\tilde{B}(s-s_0)\Big(4\mathrm{a}+\frac{\mathrm{b}}{F_\pi^2}(s-12M_\pi^2)\Big)\bigg),\\
\tilde{W}_{+0}(s,t,u)&=16\pi \bar
J(s)\bigg(\!\!\Big(\frac{\tilde{B}(s-s_0)}{2}-\tilde{A}M_\eta^2\Big)\Big(\!\mathrm{a}-\frac{\mathrm{b}\,(s+4M_\pi^2)}{2F_\pi^2}\Big)+\frac{\mathrm{b}\,\tilde{B}(s-4M_\pi^2)}{12F_\pi^2}(t-u)\!\!\bigg),\\
\tilde{\mathcal{P}}_x(s,t,u)&=\hat{\tilde{A}}M_\eta^2+\hat{\tilde{B}}(s-s_0)+\tilde{C}
(s-s_0)^2+\tilde{D}\big[(t-s_0)^2+(u-s_0)^2\big].
\end{align}
Again the hatted coefficients differ from the unhatted ones by the contribution of the $O(p^4)$ order (coming from the higher order contributions and the choice of the subtraction scheme). Thus, in the fitting of data we can proceed in two ways - either we get the starting value of the unhatted coefficients from the fit of the lower order and in the higher order allow only small variations of them and thereby actually fit only the polynomial part (eventually together with the $\pi\pi$ scattering constants); or we can perform a resummation so that we replace the unhatted coefficients by their higher order (hatted) values (making an error only of the neglected order -- here $O(p^6)$) and fit just this one set of parameters.

\subsection{Second iteration: two-loop result}
Performing the second iteration a few complications occur. In order to get the imaginary parts (in fact discontinuities) appearing in the relation (\ref{rekonstrukce}), we need to continue analytically the unitarity relation and thereby also the S and P partial wave projections of the NLO amplitudes below their physical threshold. This continuation has to be done in correspondence with the right analytic structure of the appropriate physical amplitude. This can be obtained by a careful deformation of the integration contours in the formulas for the partial wave projections (let us remind that this complication is connected with the instability of the particle $P$ and can be understood as an analytic continuation in $M_P$ going from the stable values to the instable ones). In the isospin limit this has be proven already in \cite{Bronzan Anisovich}. Their simple prescription can be used also in the case of $P^0\pi^0\rightarrow\pi\pi$ beyond the isospin limit -- we can avoid intersection of these contours with the cuts attached to normal threshold branching points of the computed amplitude and for them there also appears no anomalous threshold on the physical sheet. Nevertheless, we believe that the correct analytic continuation can be done also for the remaining processes (this will be discussed in our forthcoming papers).

A further problem is related to the fact that the analytic continuation of the discontinuities can be complex below the physical threshold (in opposition to the everywhere real discontinuity, which equals to double of imaginary part, in the simple case). As a consequence of this and also from the contribution of the sunset diagrams (see \cite{BG}), the coefficients of the polynomial $\mathcal{P}(s,t,u)$ can obtain small imaginary parts. This can be solved again in two ways sketched at the end of the previous subsection (by fitting also this small imaginary part of the hatted coefficients; or by deviation from the power counting by neglecting this imaginary part).

In the end we have the two-loop result of the $\eta\pi^0\rightarrow\pi^+\pi^-$ amplitude in the form (\ref{vysledek}) with an $O(p^8)$ error and
\begin{align}
\tilde{W}_x(s,t,u)&=\frac{1}{16\pi^2}\sum_{i=1}^5 \Big(P_i^x(s)+\frac{1}{s}Q_i^x(s)\Big)\tilde{\mathcal{G}}_i(s),\\
\begin{split}
\tilde{W}_{+0}(s,t,u)&=\frac{1}{16\pi^2}\sum_{i=1}^5 \Big(P_i^{+0}(s)+\frac{1}{s}Q_i^{+0}(s)\Big)\tilde{\mathcal{G}}_i(s)\\
&+\frac{(t-u)}{16\pi^2}\bigg(\sum_{i=1}^2 \Big(P^{+0;P}_i(s)+\frac{1}{s}Q^{+0;P}_i(s)\Big)\tilde{\mathcal{G}}_i(s)
+\sum_{i=1}^3 \Big(P^{+0;(\!\sigma\!)}_i(s)+\frac{1}{s}Q^{+0;(\!\sigma\!)}_i(s)\Big)\tilde{\mathcal{G}}^{(\!\sigma\!)}_i(s)\\
&\qquad\qquad+\sum_{i=1,2,4,5}\Big(P^{+0;(\!\lambda\!)}_i(s)+\frac{1}{s}Q^{+0;(\!\lambda\!)}_i(s)\Big)\tilde{\mathcal{G}}^{(\!\lambda\!)}_i(s)
\bigg),
\end{split}\\
\begin{split}
\tilde{\mathcal{P}}_x(s,t,u)&=\tilde{A}M_\eta^2+\tilde{B}(s-s_0)+\tilde{C}
(s-s_0)^2+\tilde{D}\left[(t-s_0)^2+(u-s_0)^2\right]
+\tilde{E}
(s-s_0)^3\\&+\tilde{F}\left[(t-s_0)^3+(u-s_0)^3\right].
\end{split}
\end{align}
For the purpose of this proceedings we do not give the explicit form of the polynomials $P_i(s)$ and $Q_i(s)$ containing the coefficients $\tilde{A}$, $\tilde{B}$, $\tilde{C}$ and $\tilde{D}$ and $\mathrm{a}$, $\mathrm{b}$, $\lambda^{(1)}$ and $\lambda^{(2)}$. The twelve $\tilde{\mathcal{G}}_i(s)$ functions are two-loop analogues of the one-loop $\bar{J}(s)$ functions (actually, $\frac{1}{16\pi^2}\tilde{\mathcal{G}}_1(s)=\bar{J}(s)$) and of the two-loop functions ${\bar K}_i(s)$ defined in the second paper in \cite{SSFKMF}.

\section{Conclusion}
We have presented a method that allows to construct a model-independent two-loop parametrization of the $P\rightarrow3\pi$ decay amplitudes based only on unitarity, analyticity, crossing symmetry, relativistic invariance and chiral power-counting for partial wave amplitudes. It contains (polynomial) subthreshold parameters describing the energy dependence of these decays and the physical parameters of $\pi\pi$ scattering (which can be either obtained from the fit of these processes or one can use their values from elsewhere to simplify the fit of the $P\pi$ polynomial parameters). We have presented the explicit result of this construction for the $\eta\rightarrow3\pi$ processes in the first order in isospin symmetry breaking.


\begin{thebibliography}{99}
\bibitem{etacharged}
A.~Abele {\it et al.},
Phys.\ Lett.\ B {\bf 417} (1998) 197;\\
%
F.~Ambrosino {\it et al.},
J.\ High Energy Phys.\ {\bf 0805} (2008) 006
[arXiv:0801.2642 [hep-ex]].
%

\bibitem{etaneutral}
A.~Abele {\it et al.},
Phys.\ Lett.\ B {\bf 417} (1998) 193;\\
%
M.~N.~Achasov {\it et al.},
JETP Lett.\ {\bf 73} (2001) 451;\\
%
W.~B.~Tippens {\it et al.},
Phys.\ Rev.\ Lett.\ {\bf 87} (2001) 192001;\\
%
F.~Ambrosino {\it et al.},
arXiv:0707.4137 [hep-ex];\\
%
M.~Bashkanov {\it et al.},
Phys.\ Rev.\ C {\bf 76} (2007) 048201
[arXiv:0708.2014 [nucl-ex]];\\
%
C.~Adolph {\it et al.},
arXiv:0811.2763 [nucl-ex];\\
%
S.~Prakhov {\it et al.},
Phys.\ Rev.\ C {\bf 79} (2009) 035204
[arXiv:0812.1999 [hep-ex]];\\
%
M.~Unverzagt {\it et al.},
Eur.\ Phys.\ J.\ A {\bf 39} (2009) 169
[arXiv:0812.3324 [hep-ex]].
%

\bibitem{kaon}
J.~R.~Batley {\it et al.} [NA48/2 Collaboration],
Phys.\ Lett.\ B {\bf 633} (2006) 173
[arXiv:hep-ex/0511056];\\
NA48 talks [see
http://na48.web.cern.ch/NA48/Welcome/images/talks.html];\\
%
E.~Abouzaid {\it et al.} [KTeV Collaboration],
Phys.\ Rev.\ D {\bf 78} (2008) 032009
[arXiv:0806.3535 [hep-ex]].

\bibitem{kaontheory}
J.~Bijnens and F.~Borg,
Nucl.\ Phys.\ B \textbf{697} (2004) 319 [arXiv:hep-ph/0405025];\\
%
N.~Cabibbo,
Phys.\ Rev.\ Lett.\ \textbf{93} (2004) 121801;\\
%
N.~Cabibbo and G.~Isidori,
J.\ High Energy Phys.\ \textbf{0503} (2005) 021 [arXiv:hep-ph/0502130];\\
%
E.~Gamiz, J.~Prades and I.~Scimemi,
Eur.\ Phys.\ J.\ C \textbf{50} (2007) 405 [arXiv:hep-ph/0602023].
%
\bibitem{Bern}
G.~Colangelo, J.~Gasser, B.~Kubis and A.~Rusetsky,
Phys.\ Lett.\ B \textbf{638} (2006) 187 [arXiv:hep-ph/0604084];\\
%
M.~Bissegger {\it et al.},
Phys.\ Lett.\ B \textbf{659} (2008) 576
[arXiv:0710.4456 [hep-ph]]
and
Nucl.\ Phys.\ B {\bf 806} (2009) 178
[arXiv:0807.0515 [hep-ph]].

\bibitem{GL}
J.~Gasser and H.~Leutwyler,
Nucl.\ Phys.\ B {\bf 250} (1985) 539.

\bibitem{BG}
J.~Bijnens and K.~Ghorbani,
J.\ High Energy Phys.\ {\bf 0711} (2007) 030
[arXiv:0709.0230 [hep-ph]].

\bibitem{KWW}
J.~Kambor, C.~Wiesendanger and D.~Wyler,
Nucl.\ Phys.\ B {\bf 465} (1996) 215
[hep-ph/9509374].

\bibitem{AL}
A.~V.~Anisovich and H.~Leutwyler,
Phys.\ Lett.\ B {\bf 375} (1996) 335
[hep-ph/9601237].

\bibitem{BN}
B.~Borasoy and R.~Nissler,
Eur.\ Phys.\ J.\ A {\bf 26} (2005) 383
[hep-ph/0510384].

\bibitem{Bijnens}
  J.~Bijnens,
  PoS {(\bf EFT09)} 022
  [arXiv:0904.3713 [hep-ph]].

\bibitem{plan} K.~Kampf, M.~Knecht, J.~Novotny, M.~Zdrahal: {\it in preparation.}

\bibitem{proceed}
  K.~Kampf, M.~Knecht, J.~Novotny and M.~Zdrahal,
  Nucl.\ Phys.\ Proc.\ Suppl.\ {\bf 186}, 334 (2009)
  [arXiv:0810.1906 [hep-ph]].

\bibitem{SSFKMF} J.~Stern, H.~Sazdjian and N.~H.~Fuchs,
Phys.\ Rev.\ D \textbf{47} (1993) 3814 [arXiv:hep-ph/9301244];\\
%
M.~Knecht, B.~Moussallam, J.~Stern and N.~H.~Fuchs,
Nucl.\ Phys.\ B {\bf 457} (1995) 513.

\bibitem{ZN}
  M.~Zdrahal and J.~Novotny,
  Phys.\ Rev.\ D {\bf 78}, 116016 (2008)
  [arXiv:0806.4529 [hep-ph]].

\bibitem{GL1}
  J.~Gasser and H.~Leutwyler,
  Annals Phys.\ {\bf 158} (1984) 142.

\bibitem{Bronzan Anisovich}
  J.~B.~Bronzan and C.~Kacser,
Phys.\ Rev.\ \textbf{132} (1963) 2703;\\
  V.~V.~Anisovich and A.~A.~Anselm,
Usp.\ Fyz.\ Nauk \textbf{88} (1966) 287.


\end{thebibliography}
\end{document}